\documentclass[prd,preprint,superscriptaddress,showpacs,byrevtex]{revtex4}
\newcommand{\be}{\begin{equation}}
\newcommand{\ee}{\end{equation}}
\newcommand{\bea}{\begin{eqnarray}}
\newcommand{\eea}{\end{eqnarray}}
\usepackage[dvips]{graphicx}
\begin{document}
%
%\draft
%
%\tighten
%\firstfigfalse
%\twocolumn[\hsize\textwidth\columnwidth\hsize\csname@twocolumnfalse\endcsname
%
\title{Color Octet Contribution to
High $p_T$ ~J/{$\Psi$} Production in pp Collisions 
at $\sqrt s$ = 500 and 200 GeV at RHIC}
\author{Gouranga C. Nayak}
\affiliation{T-8, Theoretical Division, Los Alamos National Laboratory,
Los Alamos, NM 87545, USA }
\author{Ming X. Liu}
\affiliation{P-25, Physics Division, Los Alamos National Laboratory,
Los Alamos, NM 87545, USA}
\author{Fred Cooper}
\affiliation{T-8, Theoretical Division, Los Alamos National Laboratory,
Los Alamos, NM 87545, USA}
\date{\today}
\begin{abstract}
We compute $\frac{d\sigma}{dp_T}$ of the $J/\psi$ production
in pp collisions at RHIC at $\sqrt s$ =
500 and 200 GeV by using both the color octet and singlet models in the framework
of non-relativistic QCD. The $J/\psi$ we compute here includes
the direct $J/\psi$ from the partonic fusion processes and the $J/\psi$
coming from the radiative decays of $\chi_J$'s both in the color
octet and singlet channel. The high $p_T$ $J/\psi$ production cross
section is computed within the PHENIX detector acceptance ranges:
$-0.35 ~< ~\eta~<~0.35$ and $ 1.2 ~< ~|\eta|~<~2.4$,
the central electron and forward muon arms.
It is found that the color octet contribution to $J/\psi$ production is dominant
at RHIC energy in comparison to the color singlet contributions.
We compare our results with the recent preliminary
data obtained by PHENIX detector for the high $p_T$
$J/\psi$ measurements.
While the color singlet model fails to explain the data completely
the color octet model is in agreement with the single data point above 2 GeV transverse 
momentum.
A measurement of $J/\psi$ production at
RHIC in the next run with better statistics will
allow us to determine the validity of the color octet
model of $J/\psi$ production at RHIC energies.
This is very important because it is necessary to know
the exact mechanism for $J/\psi$ production in pp collisions at RHIC
if one is to make predictions of $J/\psi$ suppression as a
signature of quark-gluon plasma.
These mechanisms also play an important role in determining the
polarized spin structure function of the proton at RHIC.

\end{abstract}
%
%\pacs{PACS: 12.38.-t, 12.38.Cy, 12.38.Mh, 11.10.Wx}
%
\maketitle
%\narrowtext
Recently there has been much progress in understanding of heavy quarkonium
production mechanism in high energy hadronic collisions. The two prominent
models for heavy quarkonium production mechanism which are quite successful at
different collider energies are 1) color singlet model \cite{sing} and 2) color
octet model \cite{oct,cho1,cho2,oct2}. 
The $J/\psi$ production measurement by CDF collaboration at
Tevatron \cite{teva} has ruled out the color singlet model for high
$p_T$ quarkonium production \cite{braaten} in that center of mass energy regime.
The color singlet model gives a reasonable description of large $p_T$
$J/\psi$ production at
ISR energies \cite{isr}. In this context large $p_T$ $J/\psi$ production
data from pp collisions at $\sqrt s$ = 500 and 200 GeV at
RHIC might tell us the relative importance and validity of these two models in
this new energy range. In particular a valid $J/\psi$
production mechanism at pp collisions will play an important role
in determining the mechanism for $J/\psi$ suppression at Au-Au collisions
which is suggested to be a prominent signature for
quark-gluon plasma detection at RHIC \cite{nayak,nayk}. These mechanisms
also play an important role in determining the polarized spin structure
function of the proton.

In the color singlet model the quarkonium is formed as a non relativistic
bound state of heavy quark-antiquark pair via static gluon exchange.
In this model it is assumed that the Q$\bar Q$ is produced
in the color singlet state at the production point with appropriate
spin (S) and orbital angular momentum
quantum number (L) to evolve to a bound state
$^{2S+1}L_J$ with total angular momentum quantum number (J).
The relative momentum of the $Q\bar Q$ pair inside
the quarkonium is assumed to be small compared to the mass $M_Q$ of
the heavy quark so that the $Q$ and $\bar Q$ will not fly apart
to form heavy mesons. The non-relativistic non-perturbative wave functions
and its derivatives appearing in the color singlet model calculation
are either determined from the potential model or
taken from experiment.

On the other hand in the color octet model relativistic
effects are taken into account which are neglected in the color singlet model.
In the color octet model using
an effective field theory called non-relativistic QCD (NRQCD), the dynamical gluon enters
into Fock state decompositions of the quarkonium states.
In NRQCD the expansion is carried out in terms of the relative velocity
$v$ ($v^2 \sim$ 0.23 for C$\bar C$ system and 0.1 for B$\bar B$ system)
of the Q$\bar Q$ bound state. The NRQCD Lagrangian density is given by:
\be
{\cal L}_{NRQCD}= {\cal L}_{light} + {\cal L}_{heavy} + {\cal L}_{correction}
\ee
where
\be
{\cal L}_{light} = -\frac{1}{4} F^{a \mu \nu}
F^a_{\mu \nu} + \sum \bar q \gamma_{\mu} D^{\mu}[A] q,
\ee
is the usual QCD Lagrangian for gluons and the light flavors and
\bea
{\cal L}_{heavy} =\psi^{\dagger} (i D_t+\frac{D^2}{2M}) \psi
+\phi^{\dagger} (i D_t+\frac{D^2}{2M}) \phi
\eea
is the leading order heavy quark part
and
\bea
{\cal L}_{correction} = \frac{1}{8M^3}
[\psi^{\dagger} D^4 \psi -\phi^{\dagger} D^4 \phi]  \nonumber \\
+\frac{g}{8M^2} [\psi^{\dagger}(D \cdot E - E \cdot D) \psi
+\phi^{\dagger}(D \cdot E - E \cdot D) \phi] \nonumber \\
+\frac{ig}{8M^2} [\psi^{\dagger} \sigma \cdot (D \times E-E \times D) \psi
+\phi^{\dagger} \sigma \cdot (D \times E-E \times D) \phi ] \nonumber \\
+\frac{g}{2M} [\psi^{\dagger} \sigma \cdot B \psi
-\phi^{\dagger} \sigma \cdot B \phi ]+...
\eea
is the correction term.
In the above equations $\psi , \phi$ are the two component Dirac spinors
of the heavy quark.
The  various correction
terms in the above equation come from the corrections to kinetic energy,
electric field coupling, magnetic field coupling and from
double electric scattering etc..
In NRQCD the dynamical gluon enter into the Fock state decompositions
of different physical states. The wave functions of
S-wave orthoquarkonia state $|\psi_Q>$ appear as:
\bea
|\psi_Q> = O(1) |Q\bar Q[^3S^{(1)}_1]>+ O(v) |Q\bar Q[^3P^{(8)}_J]g>+
O(v^2) |Q\bar Q[^3S^{(1,8)}_1]gg>+ \nonumber \\
O(v^2) |Q\bar Q[^1S^{(8)}_0]g>+
O(v^2) |Q\bar Q[^3D^{(1,8)}_J]gg>+.....
\label{jp}
\eea
and the wave functions of P-wave orthoquarkonium state $|\chi_{QJ}>$:
\bea
|\chi_{QJ}> = O(1) |Q\bar Q[^3P^{(1)}_J]>+ O(v) |Q\bar Q[^3S^{(8)}_1]g>+....
\label{cj}
\eea
In the above equation (1,8) refers to singlet and octet state of the 
$Q\bar Q$ pair.
After a Q$\bar Q$ is formed in its color octet state
it may absorb a soft gluon and transform into
$|\chi_{QJ}>$ via eq. (\ref{cj}) and then become $J/\psi$ by photon decay
or the Q$\bar Q$ in the color octet state can
emit two long wavelength gluons
and become a $J/\psi$ via eq. (\ref{jp}) and so on.
All these low energy interactions are negligible and the non-perturbative
matrix elements
can be fitted from the experiments or can be determined
from the lattice calculations.

\begin{figure}
   \centering
   \includegraphics{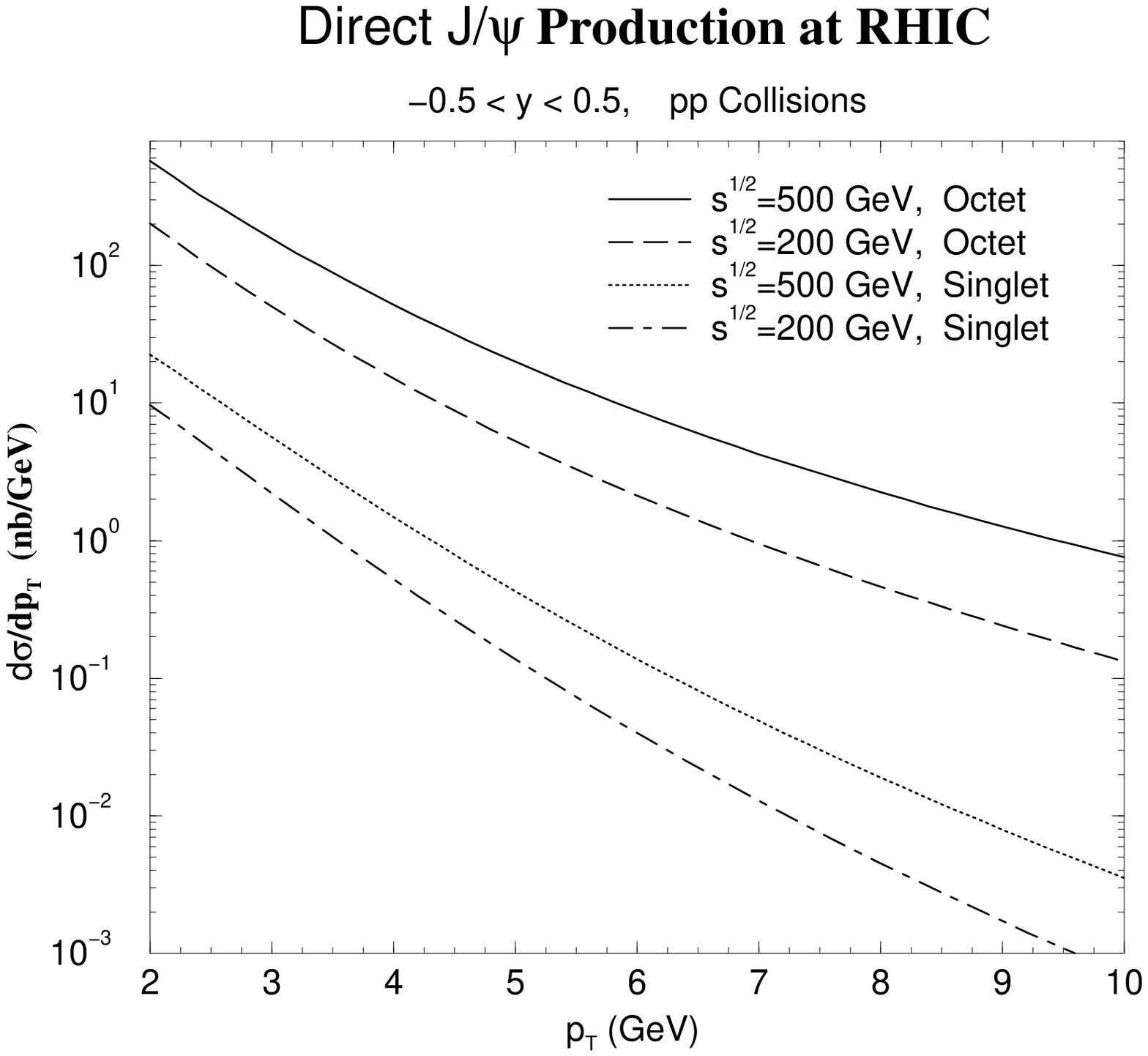}
   \caption{Direct $J/\psi$ production
 at RHIC in the $J/\psi$ rapidity range $-0.5<y<0.5$}.
\label{fig:1}
\end{figure}

Using both mechanisms for heavy quarkonium production
we compute the $p_T$ distribution of
J/$\psi$ production at RHIC
energy in this paper. The $J/\psi$ production includes
both direct $J/\psi$ and the $J/\psi$ coming from the radiative decay
of $\chi_J$'s: $\chi_J \rightarrow J/\psi + \gamma$.
We will calculate the $J/\psi$ production in this paper by considering
all the parton fusion processes.
In the color octet model the non-perturbative
matrix elements in the parton fusion processes are extracted in \cite{cho2}
by fitting CDF experimental data.

The PHENIX detector at RHIC has measured a few data points in the $p_T$ distribution
of $J/\psi$ production at $\sqrt s$ = 200 GeV pp collisions. As these data are
preliminary our comparison to this experimental data is not very
conclusive. However, in the future the PHENIX detector at RHIC will
measure high $p_T$ ~$J/\psi$ production up to $p_T$ $\sim$ 10 GeV
with better statistics. The acceptance range of single electron
for the PHENIX detector is $-0.35~<~\eta~<~0.35$ and for single
muon it is $1.2~<~|\eta|~<~2.4$. We compute
the transverse momentum distribution of $J/\psi$ production by using
both the color octet and the singlet model in the PHENIX detector acceptance range.

\begin{figure}
   \centering
   \includegraphics{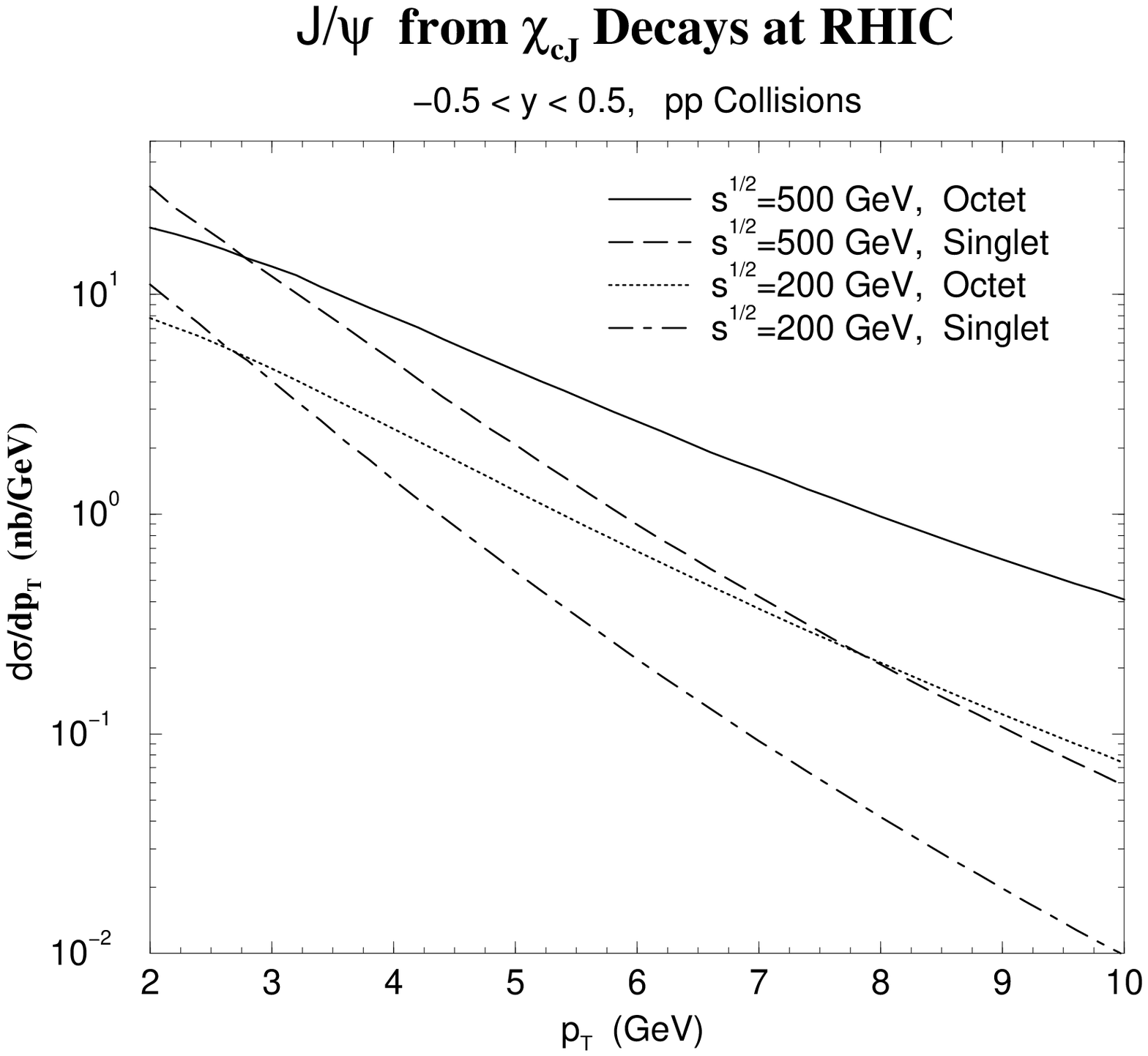}
   \caption{ $J/\psi$ production cross section at RHIC
from the $\chi_{cJ}$'s decay in the $J/\psi$ rapidity range $-0.5<y<0.5$}.
\label{fig:2}
\end{figure}

The $2 \rightarrow 2$ parton fusion processes which contribute to the
$p_T$ distribution of the $J/\psi$ and $\chi_{cJ}$ production are order  $\alpha_s^3$.
The differential cross section for the production of a quarkonium state
($^{2S+1}L_J$) with different quantum numbers (L,S,J) in the color singlet model
is given by \cite{sing}:
\bea
E\frac{d^3\sigma}{d^3p}(AB\rightarrow ^{2S+1}L_J+X)~&&
=~\sum_{i,j}\int dx_i ~x_if_{i/A}(x_i, Q^2)~\int dx_j ~x_j
f_{j/B}(x_j, Q^2)~  \nonumber \\ 
&& \times \frac{\hat{s}}{\pi}~\frac{d \hat{\sigma}}{{d\hat{t}}}(ij
\rightarrow (^{2S+1}L_J)k)   ~\delta(\hat{s}+\hat{t}+\hat{u}-M^2).
\label{dsdp}
\eea
Here $x_i$ and $x_j$ are the momentum fractions carried by partons $i$ and $j$
inside the hadrons A and B respectively.
The $f_{i/A}(x_i)$ and $f_{j/B}(x_j)$ are the parton distribution function
inside the hadron.
$\frac{d \hat{\sigma}}{{d\hat{t}}}(ij \rightarrow (^{2S+1}L_J)k)$ is the
partonic level differential cross section to form a quarkonium state
($^{2S+1}L_J$) which involves the radial wave function and its derivatives
at the origin. Using the delta function integration
in Eq. (\ref{dsdp}) one obtains:

\bea
\frac{d\sigma}{dp_T }(AB\rightarrow J/\psi, \chi_J + X)~
=~\sum_{i,j} \int dy \int dx_i ~x_if_{i/A}(x_i, Q^2)~
x_j^{\prime} f_{j/B}(x_j^{\prime}, Q^2)~  \nonumber \\
\frac{2 p_T}{{x_i -\frac{M_T}{\sqrt{s}} e^{y}}}
~\frac{d \hat{\sigma}}{{d\hat{t}}}(ij
\rightarrow ^{2S+1}L_J k),
\label{dsdp1}
\eea
where
\be
x_j^{\prime}=\frac{1}{\sqrt{s}} \frac{x_j \sqrt{s}M_T
e^{-y}-M^2}{x_j \sqrt{s}-M_T e^{y}}.
\label{x2f}
\ee
In the above $M$ is the mass of the bound state quarkonium
and $M_T^2=\sqrt{p_T^2+M^2}$. The partonic level differential
cross section
$\frac{d \hat{\sigma}}{{d\hat{t}}}(ij \rightarrow (^{2S+1}L_J)k)$
for the gluon fusion processes are obtained in \cite{gastmans} by
using helicity decomposition method and the quark, gluon
processes are derived in \cite{sing}. The
$\frac{d \hat{\sigma}}{{d\hat{t}}}(ij \rightarrow (^{2S+1}L_J)k)$
as given in \cite{sing} and \cite{gastmans} contain the non-relativistic
wave function $|R(0)|^2$ (for direct $J/\psi$ process)
and its derivatives $|R^\prime(0)|^2$ (for $\chi_J$ processes) at the
origin. For the non relativistic wave functions at the origin we take
the Buchmuller-Type wave function with charm quark mass $M_c$=1.48 GeV.
The numerical value is \cite{bm}: $|R(0)|^2$=0.81 GeV$^3$.
For the derivative of the radial wave function at origin we use
\cite{brat}
$\frac{9}{2\pi}\frac{|R^\prime(0)|^2}{M_c^4}$=15 MeV.

\begin{figure}
   \centering
   \includegraphics{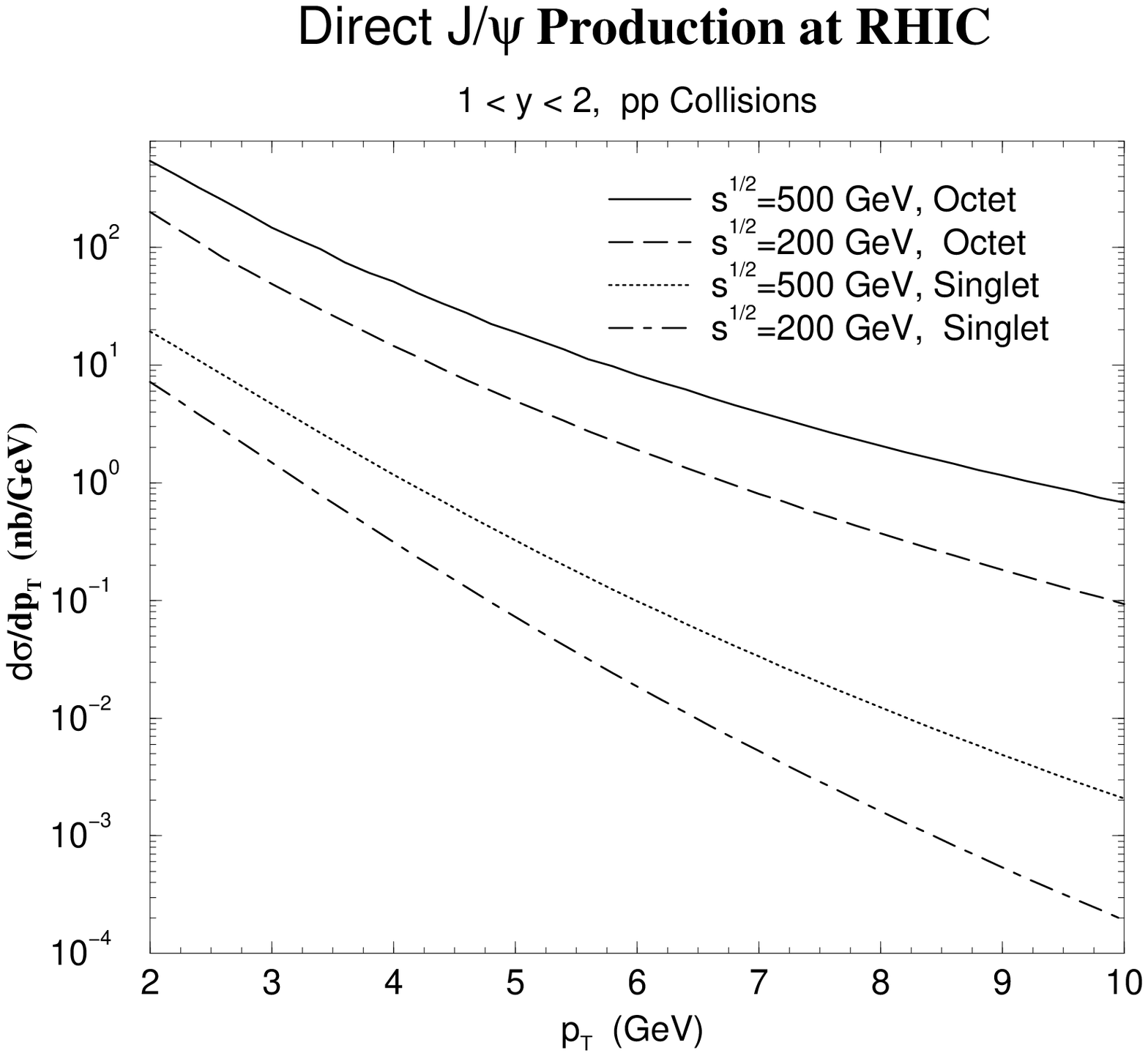}
   \caption{Direct $J/\psi$ production
 at RHIC in the $J/\psi$ rapidity range $1<y<2$}.
\label{fig:3}
\end{figure}

In the color octet model the differential cross section for heavy
quarkonium production at  order $\alpha_s^3$ is given by:
\bea
&&\frac{d\sigma}{dp_T }(AB\rightarrow \psi_Q (\chi_J) +X)~
=~\sum_{i,j} \int dy \int dx_i ~x_if_{i/A}(x_i, Q^2)~
x_j^{\prime} f_{j/B}(x_j^{\prime}, Q^2)~  \nonumber \\
&& \times \frac{2 p_T}{{x_i -\frac{M_T}{\sqrt{s}} e^{y}}}
~\frac{d \hat{\sigma}}{{d\hat{t}}}(ij \rightarrow C\bar C[^{2S+1}L_J^{(8)}]k
\rightarrow \psi_Q (\chi_J)),
\label{dsdp2}
\eea
where
\bea
&&\frac{d \hat{\sigma}}{{d\hat{t}}}(ij \rightarrow C\bar C[^{2S+1}L_J^{(8)}]k
\rightarrow \psi_Q (\chi_J))  ~=\nonumber \\
&&\frac{1}{16 \pi {\hat s}^2} \Sigma ~|{\cal{A}}
(ij \rightarrow C\bar C[^{2S+1}L_J^{(8)}]k)_{{\rm short}}|^2
<0|{\cal{O}}_8^{\psi_Q (\chi_J)}(^{2S+1}L_J)|0>.
\eea

In this paper we include the contribution from the $^3P^{(8)}_J$
and $^1S^{(8)}_0$ processes which are present in the Fock state
decomposition amplitude
(\ref{jp}) in addition to $^3S^{(8)}_1$ state of the color octet amplitude.
Similar process is followed for the $\chi_J$ production by using
(\ref{cj}) which takes into account the color octet
$^3S_1^{(8)}$ state. These $\chi_J$ formed in the color octet state
again decay radiatively to give $J/\psi$. The total
contribution to $J/\psi$ production in the color octet model
is the sum of the direct
$J/\psi$ and the contribution coming from $\chi_J$'s decay in the
color octet model.
In the above equation the short distance part of the
partonic matrix element square in different quantum states for various
partonic processes:
$\Sigma |{\cal{A}}
(ij \rightarrow C\bar C[^{2S+1}L_J^{(8)}]k)_{{\rm short}}|^2$ are very
lengthy and are given in
\cite{cho2}. We use all the partonic processes in our calculation.
We use the non-perturbative matrix elements:
$<0|{\cal{O}}_8^{\psi_Q (\chi_J)}(^{2S+1}L_J)|0>$ from the tabulation
given in \cite{cho2} which are extracted from the CDF data.
In the extraction of the these matrix elements 
$<0|{\cal{O}}_8^{\psi_Q }(^{3}P_0)|0>~=~
M_Q^2 <0|{\cal{O}}_8^{\psi_Q }(^{1}S_0)|0>$ was set and then the values
of the linear combination was extracted which is given by:
$\frac{<0|{\cal{O}}_8^{J/\psi}(^{3}P_0)|0>}{M_c^2}~+~
\frac{<0|{\cal{O}}_8^{J/\psi}(^{1}S_0)|0>}{3}~=~(2.2 \pm 0.5) \times
10^{-2}$ GeV$^3$. For other matrix elements the values are given by:
$<0|{\cal{O}}_8^{J/\psi}(^{3}S_1)|0>~=~(6.6 \pm 2.1) 10^{-3}$ GeV$^3$,
$<0|{\cal{O}}_8^{\chi_1}(^{3}S_1)|0>~=~(9.8 \pm 1.3) 10^{-3}$ GeV$^3$.
We use the central values of these matrix elements in our calculation. 

We present our results for the rapidity range covered by the central
electron arm and forward muon arm at the PHENIX detector at RHIC.
The electron pseudo rapidity range: $-0.35 <~\eta~<0.35$ corresponds to
the $J/\psi$ rapidity range: $-0.5<~y~<0.5$. For the muon pseudo rapidity
range: $1.2< ~|\eta|~<2.4$ we present our results in the $J/\psi$
rapidity range $1<~y~<2$ which is well covered by the RHIC/PHENIX experiment.
So the results we will present in this
paper is in the $J/\psi$ rapidity range: $-0.5 ~<y<~0.5$ and $1 ~<y<~2$.
We will use GRV98 NLO \cite{grv98} parton distribution function
throughout our calculation. The $Q$ scale in the structure function
and the renormalization
scale are chosen to be $Q~=~M_T~=~\sqrt{M^2+p_T^2}$, where $M$
and $p_T$ are the mass and the transverse momentum of the heavy quarkonium
respectively. In extracting the NRQCD matrix elements, the MRSD0 parton distribution
functions inside a nucleon was used in the color octet model calculation in \cite{cho2}. 
In this paper we are using, instead the GRV98 NLO parton distribution function for the
parton distribution functions. We have checked that there is no significant difference 
in the
results obtained by using MRSD0 and GRV98 in quantities calculated here. 

In Fig. 1 we present the direct $J/\Psi$ production cross section
both from the color singlet and color octet models at $\sqrt s$= 500 and 200
GeV pp collisions in the $J/\psi$ rapidity range $-0.5~<~y~<~0.5$.
The solid line is the color octet contribution
and dotted line is the color singlet contribution at $\sqrt s$=500 GeV.
The dashed line is the color octet contribution and dot-dashed line is
the color singlet contribution at $\sqrt s$=200 GeV.
The direct $J/\psi$ production from color octet model
at both the colliding energies
are much higher than the color singlet counterparts. In what follows, we present the
results of our computation for the $p_T$ range from 2 to 10 GeV.
At smaller
$p_T$, the color singlet and octet cross sections are corrupted
by collinear divergences as discussed in \cite{cho1,cho2}. Since the intrinsic
motion of the incident partons inside colliding protons renders the
differential cross section uncertain for $p_T \le $ 2 GeV, as discussed
in \cite{cho1,cho2} we have presented our results for $p_T \ge $ 2 GeV in this
paper. For low $p_T$ calculation one might follow the procedure 
where one might have to take intrinsic motion of the
incident parton into account. This calculation is, however,
beyond the scope of the present paper. In this paper we report the predictions
for high $p_T$ $J/\psi$ production. High $p_T$ limit $\sim$ 10 GeV corresponds to the
possible $p_T$ measurement range at PHENIX. In the very high $p_T$ limit ($\ge 
6 ~GeV$)
the $B$ decay contribution $B \rightarrow J/\psi +X$ and the gluon and charm
quark fragmentation contribution might become important. However, we have not
 included these
process here. In this paper we have computed only
the prompt $J/\psi$ production from the parton fusion processes.  These are the
same contributions that were used in \cite{cho2} at Tevatron energies to extract
NRQCD matrix elements.

\begin{figure}
   \centering
   \includegraphics{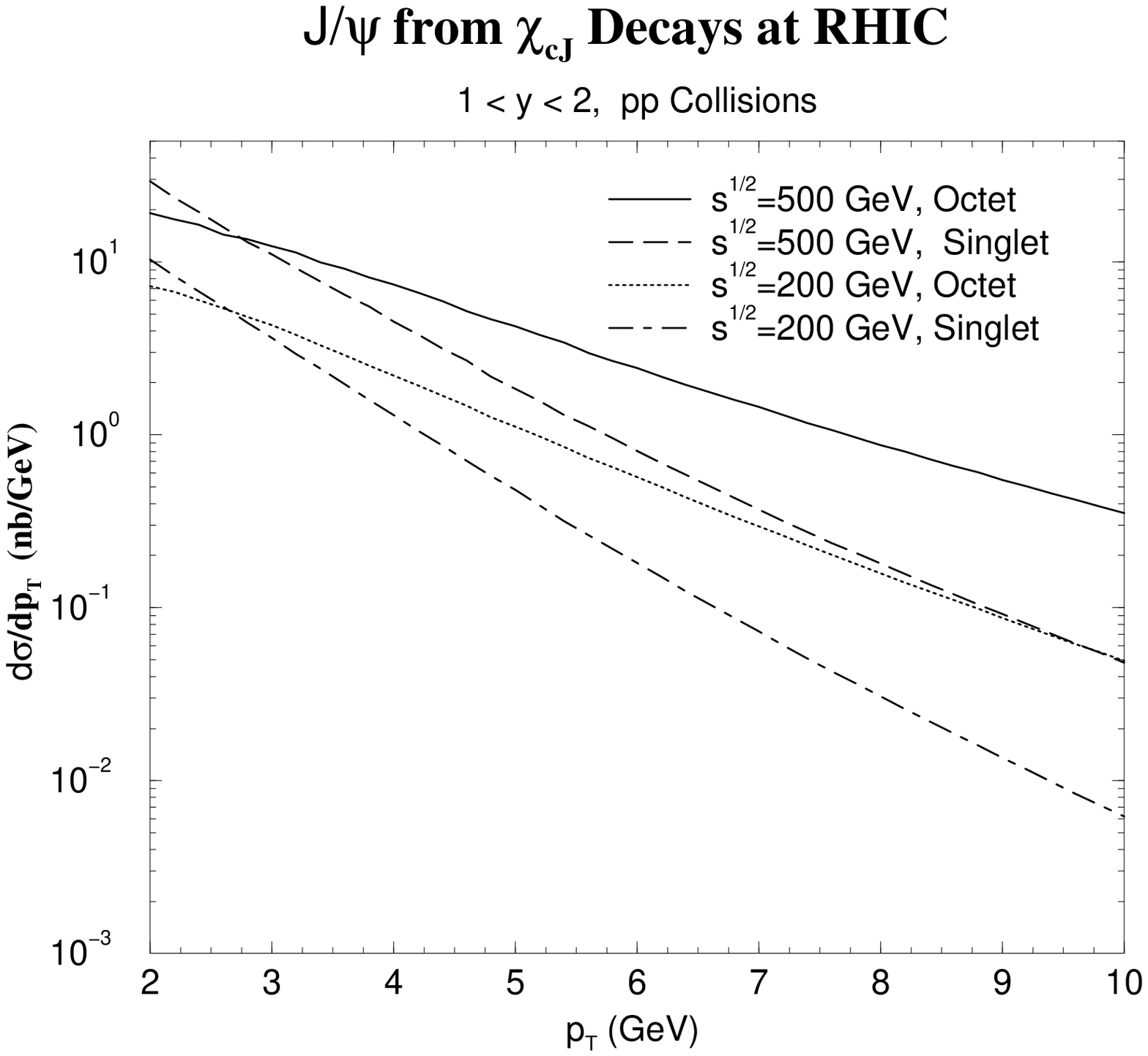}
   \caption{ $J/\psi$ production cross section at RHIC
from the $\chi_{cJ}$'s decay in the $J/\psi$ rapidity range $1<y<2$}.
\label{fig:4}
\end{figure}

In Fig. 2 we present the predictions for
 the $J/\psi$ production cross section coming from the radiative
decay of the ${\chi}_J$'s
both from the color octet and singlet model at $\sqrt s$= 500 and 200
GeV pp collisions in the $J/\psi$ rapidity range $-0.5~<~y~<~0.5$.
The solid line is the color octet contribution
and dashed line is the color singlet contribution at $\sqrt s$=500 GeV.
The dotted line is the color octet contribution and dot-dashed line is
the color singlet contribution at $\sqrt s$=200 GeV.
The $\chi_{cJ}$ contribution in each case in the figure
is the sum of the contributions from $\chi_0$, $\chi_1$ and $\chi_2$.
The $\chi_{cJ}$'s decay coming from the color singlet channel
is slightly higher at small $p_T$ but very much lower at large $p_T$
in comparison to the color octet contribution.
It can also be noted from Fig. 1 and Fig. 2 that the direct $J/\psi$
production in the color octet channel is larger than the
 contributions coming from
all other processes. Thus at RHIC energies, we find that the dominant contribution
to $J/\psi$
production comes from this direct color octet process.

In Fig.3 we present the results of our computation for the direct $J/\psi$ production
for both the color singlet and octet models in the $J/\psi$ rapidity range
$1~<~y~<~2$ at $\sqrt s$=500 and 200 GeV. The solid line is the
contribution from the color octet process and dotted line is the contribution
from the color singlet process at $\sqrt s$=500 GeV. The dashed
line and dot-dashed lines are the similar curves at $\sqrt s$=200 GeV.
In this rapidity regime we also find that the color octet contribution
 is much larger than the color singlet
contribution. 
In Fig. 4 we present the $J/\psi$ production cross section coming from
the $\chi_J$'s decay in the $J/\psi$ rapidity range $1~<y<~2$ at
$\sqrt s$= 500 and 200 GeV pp collisions at RHIC.
The solid line is the
contribution from the color octet process and dashed line is the contribution
from the color singlet process at $\sqrt s$=500 GeV. The dotted
line and dot-dashed lines are the similar curves at $\sqrt s$=200 GeV.
The total $\chi_{cJ}$ contributions in the figure includes the contribution
from $\chi_0$, $\chi_1$ and $\chi_2$.
As $\psi^{\prime}$ contribution to $J/\psi$ is small
at this energy range we have not considered the $\psi^{\prime}$ contribution
in our calculation. As can be seen again the dominant contribution to the
$J/\psi$ production comes mainly from the direct $J/\psi$ production
in the color octet channel.

The total 
$J/\psi$ $p_T$ distribution coming 
from combined color singlet model and color octet model contributions in all channels
 is plotted in Fig. 5
in the $J/\psi$ rapidity range $-0.5~<~y~<~0.5$ at $\sqrt s$=500
and 200 GeV pp collisions. The solid line is the $p_T$ distribution of the
total (octet (direct $J/\psi$ + $\chi_J$'s decay) plus singlet
(direct $J/\psi$ + $\chi_J$'s decay))
$J/\psi$ production cross section at $\sqrt s$=500 GeV. The dashed
line is the contribution from the color octet model and dot-dashed
line is the contribution from the singlet model at the same
energy. As pointed out in the above the total cross section
is the sum of the cross section from direct $J/\psi$ and from the radiative
decays of the $\chi_J$'s both in color octet and singlet channels.
It can be seen that the total cross section is almost same as the
cross section obtained from the color octet channels. The color singlet
channel contribution is very small as can be seen from the dot-dashed
line.

\begin{figure}
   \centering
   \includegraphics{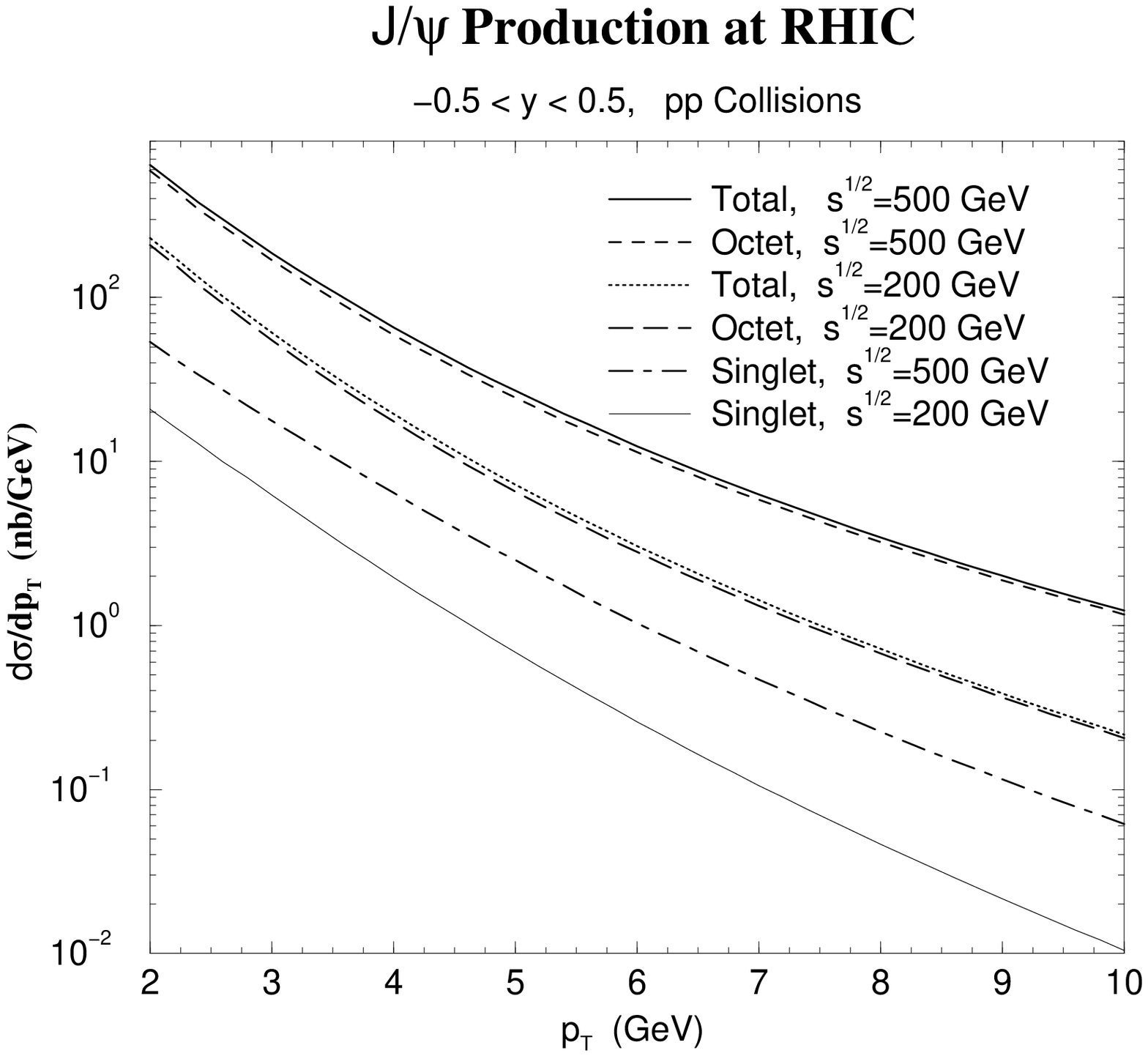}
   \caption{$J/\psi$ production at RHIC from parton fusion processes
in the $J/\psi$ rapidity range $-0.5<~y~<0.5$ }.
\label{fig:5}
\end{figure}

Similar results are obtained for $\sqrt s$=200 GeV pp collisions in the
same figure in the $J/\psi$ rapidity range $-0.5~<y<~0.5$.
The dotted line is the total
(octet (direct $J/\psi$ + $\chi_J$'s decay) plus singlet
(direct $J/\psi$ + $\chi_J$'s decay))
$J/\psi$ production cross section at $\sqrt s$=200 GeV.
Long-dashed line is the octet contribution and thin-solid line is the singlet
contribution. It can be seen from the figure that the total $J/\psi$ production cross
section is almost same as that of the color octet model contribution and
the singlet
contribution is much smaller. A measurement
of the $J/\psi$ differential cross section at RHIC in the future (both at
$\sqrt s$ = 500 and 200 GeV pp collisions) will be able to tell us whether the color
octet model reproduces the RHIC data.

\begin{figure}
   \centering
   \includegraphics{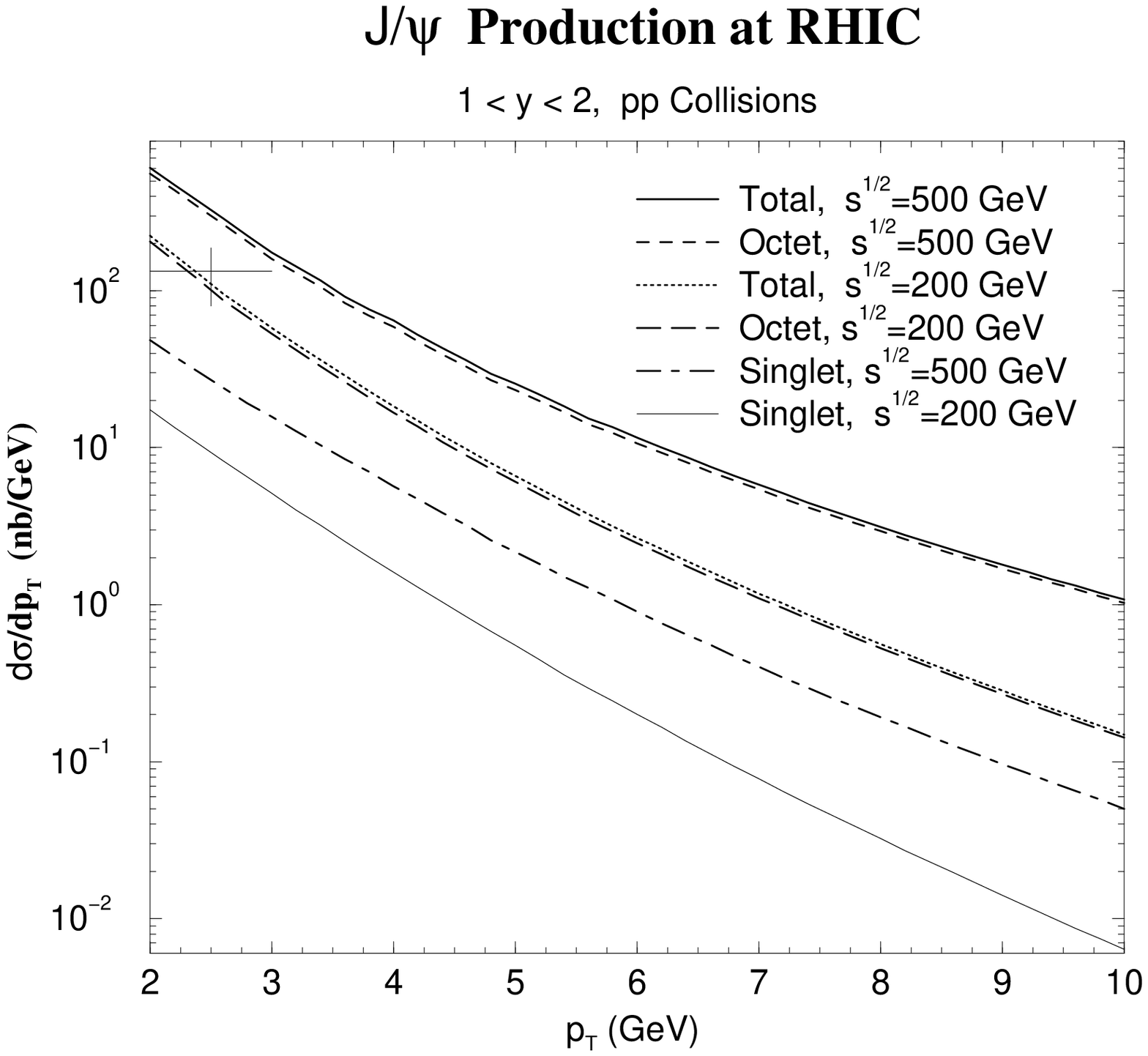}
   \caption{$J/\psi$ production at RHIC from the parton fusion processes
in the $J/\psi$ rapidity range $1<~y~<2$}.
\label{fig:6}
\end{figure}

As PHENIX has measured few data points for $p_T$ distribution of $J/\psi$
in the $\mu^+ \mu^-$ rapidity range in pp collisions at $\sqrt s$= 200 GeV,
we will compare our results with these available experimental data.
As stated earlier we will compare the available data in the
$J/\psi$ rapidity range $1<y<2$ as this range is well covered by the
RHIC/PHENIX experiment in the muon channel.
In Fig. 6 we present the $p_T$ distribution of total
$J/\psi$ production cross section
from color singlet and octet model
in the $J/\psi$ rapidity range $1~<~y~<~2$ at $\sqrt s$=500
GeV. The solid line is the $p_T$ distribution
of the
total (octet (direct $J/\psi$ + $\chi_J$'s decay) plus singlet
(direct $J/\psi$ + $\chi_J$'s decay))
$J/\psi$ production cross section at $\sqrt s$=500 GeV. The dashed
line is the contribution from the color octet model and dot-dashed
line is the contribution from the singlet model at the same
energy. The $p_T$ distribution of total
$J/\psi$ production cross section
from color octet and singlet model
in the $J/\psi$ rapidity range $1~<~y~<~2$ at $\sqrt s$=200
GeV pp collisions is plotted in Fig. 6. The dotted line is the total
(octet (direct $J/\psi$ + $\chi_J$'s decay) plus singlet
(direct $J/\psi$ + $\chi_J$'s decay))
$J/\psi$ production cross section at $\sqrt s$=200 GeV.
Long-dashed line is the octet contribution and thin-solid line is the singlet
contribution. 

Let us compare our results with the preliminary experimental data obtained at the
PHENIX detector. As mentioned earlier, at small
$p_T$ the color singlet and octet cross sections are corrupted
by collinear divergences as discussed in \cite{cho2}. Since the intrinsic
motion of the incident partons inside colliding protons renders the
differential cross section uncertain for $p_T \le $ 2 GeV, as discussed
in \cite{cho1,cho2} we will compare our results at
$p_T \ge $ 2 GeV with the PHENIX data \cite{qm02} in this
paper.  We compare
our calculation with the only available data point ($\ge$ 2 GeV)
at ~ $p_T \sim$ 2.5 GeV (see \cite{qm02}).
The data point is presented as "+" in Fig. 6. It can be seen that
the present experimental data fits the
color octet model predictions of $J/\psi$ production and
the color singlet model
prediction is well below the data point. Hence the color singlet
model  is not the relevant mechanism at RHIC energy. As for the color
octet model, the single data point matches with the color
octet contribution, but many more data with high precision measurements
are necessary to confirm the validity of the color octet
model at RHIC energy. The next run results from PHENIX experiment
will be able to shed light on this.

In summary, we have computed the $J/\psi$ production differential
cross section in pp collisions at RHIC at $\sqrt s$ =
500 and 200 GeV by using both the color octet and singlet models in the framework
of non-relativistic QCD. The $J/\psi$ production cross section
we compute here includes the direct $J/\psi$ production
from the partonic fusion processes and the $J/\psi$
coming from the radiative decays of $\chi_J$'s both in the color
octet and singlet channel. The high $p_T$ $J/\psi$ production cross
section is computed within the $J/\psi$ rapidity range:
$-0.5 ~< ~y~<~0.5$ and $ 1 ~< ~|\eta|~<~2$, in the
the central electron and forward muon arms respectively.
It is found that the color octet contribution to $J/\psi$ production is dominant
at RHIC energy in comparison to the color singlet contributions.
We have compared our results with the recent
data obtained by PHENIX detector for $p_T \ge $ 2 GeV at $\sqrt s$ =200 GeV
pp collisions.
While the color singlet model fails to explain the data completely
the color octet model seems to reproduce the data.
As the present high $p_T$ data of $J/\psi$ production
at PHENIX detector at RHIC is very restricted,
measurement of high $p_T$ $J/\psi$ production at RHIC in the
next run with better statistics will
be able to tell us the degree of validity of the color octet
model of $J/\psi$ production at RHIC energies.
It is necessary to know
the exact mechanism of $J/\psi$ production in  pp collisions at RHIC
in order to make a prediction of $J/\psi$ suppression as a
signature of the quark-gluon plasma in Au-Au collisions. Knowledge of the
$J/\psi$ production mechanism is also crucial for isolating the 
polarized spin structure functions of the proton from the data. 

\acknowledgments{}
This research is supported by the Department of Energy, under contract
W-7405-ENG-36.
We thank Rajiv Gavai, Mike Leitch, Pat McGaughey, Emil Mottola
and Johann Rafelski for useful discussions.

\end{document}